\documentclass[12pt,preprint]{aastex}

\newcommand{\eqref}[1]{(\ref{#1})}

\begin{document}

\title{Variational Integrators for the Gravitational $N$-Body Problem}
\author{Will M. Farr \and Edmund Bertschinger} 

\affil{Department of Physics and Kavli Institute for Astrophysics and
  Space Research, MIT Room 37-602A, 77 Massachusetts Ave., Cambridge,
  MA 02139}

\email{farr@mit.edu, edbert@mit.edu}

\begin{abstract}
This paper describes a fourth-order integration algorithm for the
gravitational $N$-body problem based on discrete Lagrangian mechanics.
When used with shared timesteps, the algorithm is momentum conserving
and symplectic.  We generalize the algorithm to handle individual time
steps; this introduces fifth-order errors in angular momentum
conservation and symplecticity.  We show that using adaptive block
power of two timesteps does not increase the error in symplecticity.
In contrast to other high-order, symplectic, individual timestep,
momentum-preserving algorithms, the algorithm takes only forward
timesteps.  We compare a code integrating an $N$-body system using the
algorithm with a direct-summation force calculation to standard
stellar cluster simulation codes.  We find that our algorithm has
about 1.5 orders of magnitude better symplecticity and momentum
conservation errors than standard algorithms for equivalent numbers of
force evaluations and equivalent energy conservation errors.
\end{abstract}


 \maketitle

\section{Introduction}

The gravitational $N$-body problem is the numerical integration of
trajectories for $N$ particles with pairwise inverse square-law
forces.  Ideally, a numerical method should respect all of the
symmetries of the exact problem.  Conservation of linear and angular
momentum and energy --- in toto or in pairwise interactions --- are
often used as quality indicators for numerical algorithms.  However,
there is another conservation law following from Hamiltonian dynamics:
conservation of the symplectic form on phase space.

Symplectic integration methods conserve exactly the symplectic form.
This leads to other desirable properties of the flow of the
integrator; for example, symplecticity implies incompressible, or
dissipation less, flow.  Symplectic methods are well known in
numerical integration of the solar system following the pioneering
work of \cite{Wisdom1991}.  However, the advantages of symplectic
methods apply to systems with many more bodies.  Indeed, cosmological
simulations with millions or even billions of particles often use the
leapfrog method because of its symplectic behavior
\citep{Springel2005,Shirokov2005}.

The leapfrog algorithm is second-order accurate.  Higher-order
accurate compositional algorithms are possible but conventionally
require sub-steps integrated backwards in time
\citep{Yoshida1993,Chambers2003} and such techniques typically require
many more force evaluations than standard integrators.  Symplectic
implicit Runge-Kutta-Nystr\"{o}m integrators are well-known, and involve
relatively few force evaluations
\citep{Suris1989,Marsden2001,Stuchi2002}.

Gravitating systems typically have a large dynamic range of density
and hence dynamical time, making it computationally inefficient to use
a constant timestep, as required by most symplectic algorithms.
Either adaptive timesteps (which change with time as a system
evolves), individual timesteps (which differ for each particle), or
both are required to make a computation feasible.  This is especially
true when unsoftened inverse-square law forces are used, e.g., in
numerical simulation of globular clusters \citep{HeggieHut2003}.  It
is well-known how to use compositional symplectic algorithms with
individual timesteps \citep{Springel2005}, but the larger number of
force evaluations in the high-order compositional algorithms and the
necessity of backward sub-steps rule these out for use in simulations
of large-$N$ gravitating systems \citep{SpringelPrivate2005}.  No
individual timestepping symplectic Runge-Kutta-Nystr\"{o}m algorithm has
appeared previously in the literature.

Below we describe a fourth-order symplectic Runge-Kutta-Nystr\"{o}m
algorithm from a variational point of view.  The algorithm requires
the solution of a nonlinear algebraic equation for one forwards
sub-step.  If the nonlinear equation is solved approximately by
iteration then the algorithm is approximately symplectic with an error
in phase space conservation that can be made arbitrarily small.  We
describe how to use data from the last step of the integrator to
generate an approximate solution sufficient to obtain fifth-order
symplecticity and momentum-conservation using only two force
evaluations per step.  We generalize this algorithm to use individual
(and adaptive) timesteps; the formulation in terms of a discretized
action principle is essential to the generalization.  This is the
first individual timestepping symplectic Runge-Kutta-Nystr\"{o}m
algorithm to appear in the astrophysical literature.

It is widely believed that adaptive timesteps are incompatible with
symplecticity but we show otherwise in Section
\ref{AdaptiveTimesteps}. This assumed breakdown of symplecticity,
coupled with the large number of force evaluations for standard
higher-order compositional symplectic methods and the lack of
individual timestepping symplectic Runge-Kutta-Nystr\"{o}m algorithms, has
led researchers to seek alternative ways to reduce dissipation,
e.g.\ using time-reversible integration
\citep{Makino1996,PretoTremaine1999}.  These techniques treat the
symptoms of non-symplecticity (linear growth in various errors)
without treating the cause (non-conservation of the symplectic
form). Our algorithm addresses the cause, and we see corresponding
improvements in symplecticity and momentum conservation for equivalent
energy error to relative standard algorithms.

In this paper we present a fourth-order integrator requiring only two
force evaluations per timestep, which is fifth-order in
symplecticity. Each additional force evaluation improves the
symplecticity by two powers of the timestep.  We generalize the
integrator to individual timesteps and analyze the breakdown of
symplecticity when adaptive and individual timesteps are used.  We
show that symplecticity is effectively restored when block power of
two timesteps are used.

The algorithm is based on a discrete approximation to the action of a
system, described in Section \ref{VariationalIntegrators}.  The
algorithm contains a non-linear equation whose solution must be
approximated; we compare two approximation methods in Section
\ref{IterationPrescriptions}.  Adaptive, individual, and combined
block timesteps are discussed in Sections \ref{AdaptiveTimesteps},
\ref{IndividualTimeSteps}, and \ref{Implementation} respectively.
Numerical tests are presented in Section \ref{SimulationResults}.
Conclusions are given in Section \ref{Conclusions}.

\section{Variational Integrators}
\label{VariationalIntegrators} Variational integrators are based on
applying Hamilton's principle of stationary action to discrete
approximations to the action for a physical system.  \citet{Lew2004}
is an excellent introduction to variational integrators in an
engineering context; \citet{Marsden2001} provides a much more
mathematical discussion, including proofs of the essential
properties of variational integrators and many examples of
particular integration rules.  This section is a brief introduction
to variational integrators.  Here and throughout we suppress vector
indices on variables (juxtaposition of variables thus denotes
multiplication in the one-dimensional case and the usual dot-product
in the multidimensional case).  We denote the derivative of the
function $f$ by $Df$; we denote the partial derivative on the $i$th
argument of the function $g$ by $\partial_i g$ (argument labels
begin at 0).

The fundamental theorem of variational integration \citep[Theorem
2.3.1]{Marsden2001} states that if $H$ is an approximation to the
action of a mechanical system,
\begin{equation}
H(h, q_1, q', q'', \ldots, q^{(n)}, q_2) = S[q](t_1, t_1+h) +
\mathcal{O}(h^{r+1}) = \int_{t_1}^{t_1+h} dt \, L(t, q(t), Dq(t)) +
\mathcal{O}(h^{r+1}),
\end{equation}
where $q_1 = q(t_1)$, $q_2 = q(t_2)$, the $q^{(i)}$ are intermediate
positions in the time interval $[t_1, t_1+h]$, $S$ is the action
functional and $L$ is the Lagrangian for the system, then the
equations
\begin{mathletters}
\label{evolutionEquations}
\begin{eqnarray}
\label{p1}
\partial_1 H(h, q_1, q', q'', \ldots, q^{(n)}, q_2) & = & -p_1 \\
\label{middlePoints}
\partial_i H(h, q_1, q', q'', \ldots, q^{(n)}, q_2) & = & 0, \qquad i = 2,3,\ldots,n+1 \\
\label{p2}
\partial_{n+2} H(h, q_1, q', q'', \ldots, q^{(n)}, q_2) & = & p_2
\end{eqnarray}
\end{mathletters}
define a map $(q_1, p_1) \mapsto (q_2, p_2)$ which is an order $r$
integrator for the mechanical system.  The function $H$ is called
the \emph{discrete Lagrangian}.  Equations  \eqref{middlePoints}
extremize the discrete action approximation with respect to the
discrete path $\{q^{(i)}\}$, while equations  \eqref{p1} and
\eqref{p2} exploit that the action is a $F_1$-type generating
function for the time-evolution canonical transformation
\citep[pp.~415--416]{Sussman2001}.

The map defined by equations \eqref{evolutionEquations} has many
useful properties analogous to the properties of the exact evolution
of the system defined by $L$.  First, it is momentum-preserving:
imagine an infinitesimal variation in the coordinates $q_1$, $q_2$
and $q^{(i)}$ which leaves $H$ invariant when $q_1$, the $q^{(i)}$
and $q_2$ satisfy equations \eqref{evolutionEquations}.  We have
\begin{eqnarray}
\label{momentumConservation} \delta H & = & \partial_1 H(h, q_1, q',
q'', \ldots, q^{(n)}, q_2) \delta q_1 + \sum_{i = 2}^{n+1} \partial_i
H(h, q_1, q', q'', \ldots, q^{(n)}, q_2) \delta q^{(i)} \nonumber \\ &
& \mbox{} + \partial_{n+2} H(h, q_1, q', q'', \ldots, q^{(n)}, q_2) \delta q_2
\nonumber \\ & = & p_2 \delta q_2 - p_1 \delta q_1 \nonumber \\ & = & 0.
\end{eqnarray}
In this situation the quantity $p \delta q$ is conserved by the
integrator: this is the discrete version of N\"{o}ther's theorem.
Assuming that $H$ inherits the symmetries of $L$, the integrator
will \emph{exactly} preserve the associated discrete momenta.

Second, the map is \emph{symplectic}.  Consider the discrete
approximation to the action over an interval evaluated on the
integrator path:
\begin{equation}
\mathcal{S}\left( q_1, q_2, \ldots, q_M \right) \equiv \sum_{i =
1}^{M-1} H\left( h, q_i, q'_i, \ldots, q^{(n)}_i, q_{i+1} \right),
\end{equation}
where the $q_i$ satisfy the integrator equations
\eqref{evolutionEquations}.  Taking one exterior derivative of
$\mathcal{S}$ gives
\begin{eqnarray}
\label{dS} d \mathcal{S}\left( q_1, q_2, \ldots, q_M \right) & = & 
\partial_1 H \left( h, q_1, q'_1, \ldots, q^{(n)}_1, q_2 \right) dq_1
\nonumber \\ & & \mbox{} + \partial_{n+2} H\left( h, q_{M-1}, q'_{M-1},
\ldots, q^{(n)}_{M-1}, q_M \right) dq_{M};
\end{eqnarray}
the terms involving $dq_2, \ldots, dq_{M-1}$ are identically zero
because the trajectory satisfies equations  \eqref{p1},
\eqref{middlePoints} and  \eqref{p2}.  Two exterior derivatives of
$\mathcal{S}$ give zero, yielding
\begin{eqnarray}
\label{d2S}
d^2 \mathcal{S}\left( q_1, q_2, \ldots, q_M \right) & = & \partial_1
\partial_{n+2} H \left( h, q_1, q'_1, \ldots, q^{(n)}_1, q_2 \right)
dq_1 \wedge dq_2 \nonumber \\ & & \mbox{} + \partial_1 \partial_{n+2}
H\left( h, q_{M-1}, q'_{M-1}, \ldots, q^{(n)}_{M-1}, q_M \right)
dq_{M} \wedge dq_{M-1} \nonumber \\ & = & 0.
\end{eqnarray}

Instead of considering evolution on phase space $(q,p)$, consider
the corresponding evolution on the discrete state space $\left( q_1,
q_2 \right)$.  Evolution maps the initial state-space for $H$,
$\left(q_1,q_2 \right) \in \mathbb{R}^m \times \mathbb{R}^m$, to an
isomorphic space, $\left( q_{M-1}, q_{M} \right) \in \mathbb{R}^m
\times \mathbb{R}^m$, where $m$ is the dimensionality of
configuration space.  Equation  \eqref{d2S} can be written
\begin{equation}
\label{symplecticity}
\partial_1 \partial_{n+2} H \left( h, q_1, q'_1, \ldots,
q^{(n)}_1, q_2 \right) dq_1 \wedge dq_2 = F^* \left[ \partial_1
\partial_{n+2} H\left( h, q_1, q'_1, \ldots, q^{(n)}_1, q_2 \right)
dq_{1} \wedge dq_{2} \right]
\end{equation}
using the pushforward map under evolution, $F^*$.  All forms in
equation \eqref{symplecticity} live on the cotangent bundle of the
state space $\mathbb{R}^m \times \mathbb{R}^m$.  We see that the
integrator conserves the discrete symplectic form on the state space
of $H$,
\begin{equation}
\label{symplecticForm}
\partial_1 \partial_{n+2} H \left( h, q_1, q'_1, \ldots,
q^{(n)}_1, q_2 \right) dq_1 \wedge dq_2.
\end{equation}
This is the direct analog of the symplecticity of continuous
time-evolution in a Hamiltonian system.

Using equation \eqref{p1}, we see that
\begin{equation}
- dp_1 = \partial_1 \partial_1 H(h, q_1, q', q'', \ldots, q^{(n)},
q_2) dq_1 + \partial_{n+2} \partial_1 H(h, q_1, q', q'', \ldots,
q^{(n)}, q_2) dq_2,
\end{equation}
and therefore conservation of the discrete symplectic form in
equation \eqref{symplecticForm} implies conservation of the
Poincar\'{e} integral invariant on phase space:
\begin{equation}
\label{dqWedgedpConservation}
\partial_1 \partial_{n+2} H \left( h, q_1, q'_1, \ldots,
q^{(n)}_1, q_2 \right) dq_1 \wedge dq_2 = dp_1 \wedge dq_1.
\end{equation}

Finally, while it is impossible for a constant-time-stepping
integrator to be momentum-conserving, symplectic and to exactly
conserve energy \citep{Ge1988}, variational integrators generally
have bounded energy error.  \citet{Lew2004} explain that the
discretized trajectory is sampling the continuous trajectory of a
Lagrangian system, $\tilde{L}$, which is near $L$.  $\tilde{L}$
satisfies
\begin{equation}
\label{LTilde} \int_{t_1}^{t_1+h} dt \, \tilde{L}(t, q(t), Dq(t)) =
H\left(h, q_1, q', q'', \ldots, q^{(n)}, q_2\right),
\end{equation}
where the integral is evaluated on the trajectory which satisfies
the Euler-Lagrange equations for $\tilde{L}$ with $q\left( t_1
\right) = q_1$ and $q\left( t_1 + h \right) = q_2$ and the positions
which are arguments for $H$ satisfy equations
\eqref{evolutionEquations}.  Equation \eqref{LTilde} implies that
$H$ is the exact generating function for time evolution under
$\tilde{L}$.  In general it is only possible to compute a truncation
of $\tilde{L}$ to any desired order in $h$, but it is possible to
prove that $\tilde{L}$ is close to $L$ in the space of possible
Lagrangians.  Since the trajectory remains on the energy level set
of $\tilde{L}$ in phase space, near the energy level-set for $L$,
the energy error remains bounded.

\subsection{Galerkin Gauss-Lobatto Variational Integrators}

To define a variational integrator, we need an $H$-type function
which approximates the action for a system over an interval.  There
are many ways to find such a function; see \citet{Marsden2001} for
an extensive discussion of the various types of integrator.  In this
paper, we will focus on the so-called Galerkin Gauss-Lobatto (GGL)
integrators.  These integrators assume a polynomial trajectory in
time and approximate the action integral using a Gauss-Lobatto
quadrature rule.  Gauss-Lobatto quadrature is appropriate because it
gives the highest-order integration rule for a given number of
points subject to the constraint that the Lagrangian is evaluated
once at the beginning and once at the end of the interval.
Evaluating at the beginning and end of the interval is important
because it preserves the symmetries of the continuous Lagrangian.

\citet{Marsden2001} show that all GGL integrators can be written as
symplectic partitioned Runge-Kutta integrators and derive formulas
which relate the SPRK coefficients to the discrete Lagrangian.
Because variational integrators are symplectic, GGL integrators
automatically satisfy the constraints on symplectic Runge-Kutta
coefficients (see, for example,
\citet{Suris1989,Marsden2001,Stuchi2002}).  The particular integration
algorithms we consider here belong to a sub-class of partitioned
Runge-Kutta methods often called Runge-Kutta-Nystr\"{o}m methods.

We define
\begin{eqnarray}
\label{quadH} 
\lefteqn{H(h, q_1, q', \ldots, q^{(n)}, q_2) \equiv} \nonumber \\ & &
w_1 L(t_1, \phi(t_1, q_1, q', \ldots, q^{(n)}, q_2, t_1, h),
\partial_0 \phi(t_1, q_1, q', \ldots, q^{(n)}, q_2, t_1, h)) \nonumber
\\ & & \mbox{} + \sum_{i = 1}^n w^{(i)} L(t^{(i)}, \phi(t^{(i)}, q_1,
q', \ldots, q^{(n)}, q_2, t_1, h), \partial_0 \phi(t^{(i)}, q_1, q',
\ldots, q^{(n)}, q_2, t_1, h)) \nonumber \\ & & \mbox{} + w_2 L(t_1+h,
\phi(t_1+h, q_1, q', \ldots, q^{(n)}, q_2, t_1, h), \partial_0
\phi(t_1+h, q_1, q', \ldots, q^{(n)}, q_2, t_1, h)),
\end{eqnarray}
where $\phi(t, q_1, q', \ldots, q^{(n)}, q_2, t_1, h)$ is the
interpolating polynomial for $q(t)$ passing through the points
$\{q_1, q', \ldots, q^{(n)}, q_2\}$ at times $\{t_1, t', \ldots,
t^{(n)}, t_1+h\}$.  The times $\{t_1, t', \ldots, t^{(n)}, t_1+h\}$
and weights $\{w_1, w', \ldots, w^{(n)}, w_2\}$ define the $n+2$
point Gauss-Lobatto quadrature rule.  This rule is exact for
quadrature of polynomials up to and including order $2n + 2$.  The
integrator so defined will be of order $2n+2$.  See, for example,
\citet[Table 25.6]{Abramowitz1972} for appropriate times and
weights.  This is easier than it looks; examples follow.

\subsubsection{Two-Point Integrator}
\label{2ptIntegrator} The two-point Gauss-Lobatto integration rule
has times $\{t_1, t_1+h\}$ and weights $\{h/2, h/2\}$; a two-point
interpolating polynomial is a line.  Therefore, we have
\begin{equation}
H(h, q_1, q_2) = \frac{h}{2} \left[ L\left(t_1, q_1,
\frac{q_2-q_1}{h} \right) + L\left(t_1+h, q_2, \frac{q_2-q_1}{h}
\right) \right].
\end{equation}
For a Lagrangian $L(t,q,v) = \frac{1}{2} m v^2 - V(q)$, equations
\eqref{p1} and  \eqref{p2} result in the explicit integration rule
\begin{mathletters}
\begin{eqnarray}
q_2 & = & q_1 + h \frac{p_1}{m} - \frac{h^2}{2m} DV(q_1) \\
p_2 & = & p_1 - \frac{h}{2} \left[ DV\left(q_1\right) + DV\left(q_2\right)\right].
\end{eqnarray}
\end{mathletters}
This rule is kick-drift-kick leapfrog; it can be derived from the
Hamiltonian viewpoint by iterating the evolutions of the splitting
of the Hamiltonian for this system into $H_1(q,p) = V(q)$ and
$H_2(q,p) = p^2/(2m)$ (for a thorough exploration of this idea see,
for example, \citet{Wisdom1991} and \citet{Yoshida1993}).  This rule
is second order, as expected from the order of the quadrature rule.

\subsubsection{Three-Point Integrator}
\label{ThreePointIntegrator} The three-point Gauss-Lobatto
quadrature rule has times $\{t_1, t_1+h/2, t_1+h\}$ and weights
$\{h/6, 2h/3, h/6\}$.  The three-point interpolation polynomial is
quadratic in time.  Applying equation  \eqref{quadH}, we find
\begin{eqnarray}
H(h, q_1, q', q_2) & = & h \left[ \frac{1}{6}L\left(t_1, q_1,
  \frac{4q' - 3 q_1 - q_2}{h} \right) \right. \nonumber \\ & &
  \left. \mbox{} + \frac{2}{3} L\left( t_1+\frac{h}{2}, q', \frac{q_2
    - q_1}{h} \right) + \frac{1}{6} L\left(t_1+h, q_2, \frac{q_1 + 3
    q_2 - 4 q'}{h} \right) \right].
\end{eqnarray}
If $L(t,q,v) = \frac{1}{2} m v^2 - V(q)$ then equations  \eqref{p1},
\eqref{middlePoints} and  \eqref{p2} reduce to
\begin{mathletters}
\label{3ptEquations}
\begin{eqnarray}
\label{implicit}
q' & = & q_1 + \frac{h}{2} \frac{p_1}{m} - \frac{1}{2}
\left(\frac{h}{2}\right)^2 \left[ \frac{2 DV\left(q_1\right)}{3m} +
\frac{DV\left( q' \right)}{3m} \right] \\
\label{3pt-q2}
q_2 & = & q_1 + h \frac{p_1}{m} - \frac{1}{2} h^2 \left[
\frac{DV\left(q_1\right)}{3m} + \frac{2 DV\left( q' \right)}{3m}
\right] \\
\label{3pt-p2}
p_2 & = & p_1 - h \left[ \frac{DV\left(q_1\right)}{6} + \frac{2
  DV\left( q' \right)}{3} + \frac{DV\left( q_2 \right)}{6} \right]
\end{eqnarray}
\end{mathletters}
To implement this integration scheme, we must solve the implicit
equation \eqref{implicit} for $q'$.  In spaces of high dimensionality
(i.e.~an $N$-body system with large $N$), the only efficient way to do
this is through iteration: we treat equation \eqref{implicit} as a
prescription for updating the value of $q'$, and iterate.  We describe
two iteration techniques in the following section.

The three-point GGL integrators take only forward steps in time.  It
is not possible to write these integrators as iterated evolution of a
splitting of the Hamiltonian corresponding to $L$ (as in the two-point
case): a general theorem states that it is impossible to have
evolution by Hamiltonian splitting which is higher than second-order
accurate and takes only forward steps \citep{Sheng1989,Suzuki1991}.

It is possible to formulate higher-order mapping integrators from the
Hamiltonian perspective which take only forward steps using the
\emph{force} gradient \citep{Wisdom1996,Scuro2005,Chin2003}, but force
gradients can be expensive to compute.  \citet{Omelyan2006} describes
how to approximate computation of a force gradient with an extra force
evaluation at a shifted position.  The iteration technique in Section
\ref{OmelyanIteration} reproduces algorithm 8 in \citet{Omelyan2006}
from the equations \eqref{implicit}, \eqref{3pt-q2}, and
\eqref{3pt-p2}.  However, we find that for equivalent numbers of force
evaluations in the $N$-body problem, the alternate iteration technique
in Section \ref{OurIteration} typically outperforms the one in Section
\ref{OmelyanIteration} by one to two orders of magnitude in energy
error (see Figure \ref{OmelyanVsUs}), so the gradient-approximating
algorithm from \citet{Omelyan2006} is sub-optimal for our purposes.

Forward time steps are important for cosmological simulations which
include gas dynamics because such simulations are unstable under time
reversal.  The requirement of symplecticity and forward timesteps has
previously restricted cosmological simulations to second-order mapping
integrators \citep{SpringelPrivate2005}.

It is straightforward to derive the integration equations for
$n+2$-point GGL integrators with $n > 1$.  All such integrators have
implicit equations for the intermediate positions, $q', \ldots,
q^{(n)}$, which must be solved via iteration exactly as in the $n = 1$
case discussed above.  In the presence of individual timesteps
(Section \ref{IndividualTimeSteps}), we must predict the intermediate
positions---we cannot iterate the implicit equations to convergence.
For integrators of order greater than four, such predicted positions
do not solve the implicit equations accurately enough to make the
symplecticity error scale better than the trajectory error; as we
shall see, we can predict the solution to equation \eqref{implicit}
accurately enough to produce fifth-order symplecticity error in the
fourth-order integrator.  For this reason, the fourth-order integrator
is uniquely positioned in the hierarchy of GGL integrators for
adaptation to individual timesteps

\section{Solving the Implicit Equation}
\label{IterationPrescriptions}

This section discusses two ways to solve equation \eqref{implicit}
with iteration.  They differ in their choice of initial guess for
$q'$.  The choice in Section \ref{OmelyanIteration} produces an
algorithm which is compositional, and is equivalent to algorithm 8
from \citet{Omelyan2006}.  That algorithm is exactly
phase-space-volume and momentum conserving, but only fourth-order
symplectic, with three force evaluations per step.  The choice in
Section \ref{OurIteration} produces an algorithm which is fifth-order
symplectic (and the same in phase-space-volume error), exactly
conserves linear momentum, and conserves angular momentum at
fifth-order, with two force evaluations per step.

\citet{Omelyan2006} reports excellent energy conservation for the
algorithm in Section \ref{OmelyanIteration} when simulating the
one-dimensional Kepler problem (where phase-space-volume conservation
implies symplecticity), but we find that the algorithm in Section
\ref{OurIteration} has one to two orders of magnitude better energy
conservation for equivalent numbers of force evaluations in the
$N$-body problem for $N > 2$ (see Figure \ref{OmelyanVsUs}).  This is
probably due to the superior symplecticity of the algorithm in Section
\ref{OurIteration} for multidimensional configuration spaces.  We do
not discuss the generalization of the algorithm in Section
\ref{OmelyanIteration} to individual and adaptive timesteps, but
instead focus on the algorithm in Section \ref{OurIteration} for the
remainder of the paper.

\subsection{Compositional Algorithm}
\label{OmelyanIteration}

To reproduce algorithm 8 in \citet{Omelyan2006}, let the initial guess
for $q'$ be
\begin{equation}
q'_{(0)} = q_1 + \frac{h}{2} \frac{p_1}{m} - \frac{1}{2}
\left(\frac{h}{2}\right)^2 \frac{2 DV\left(q_1\right)}{3m}.
\end{equation}
Then iterate equation \eqref{implicit} once, producing
\begin{equation}
q' = q'_{(0)} - \frac{h^2}{24m} DV\left( q'_{(0)} \right).
\end{equation}
$q_2$ and $p_2$ are then given by equations \eqref{3pt-q2} and
\eqref{3pt-p2}, with this choice for $q'$.  This choice of initial
guess and single iteration allows the algorithm to be written
compositionally as
\begin{mathletters}
\begin{eqnarray}
p & \leftarrow & p - \frac{h}{6} DV(q) \\
q & \leftarrow & q + \frac{h}{2} \frac{p}{m} \\
p & \leftarrow & p - \frac{2 h}{3} DV\left( q - \frac{h^2}{24m} DV(q)\right) 
\label{gradientApprox} \\
q & \leftarrow & q + \frac{h}{2} \frac{p}{m} \\
p & \leftarrow & p - \frac{h}{6} DV(q).
\end{eqnarray}
\end{mathletters}
This is exactly the sequence of operations in \citet{Omelyan2006},
algorithm 8; equation \eqref{gradientApprox} is the approximation
derived in \citet{Omelyan2006} to the force gradient required in the
corresponding algorithm of \citet{Chin2003}.  The algorithm needs four
force evaluations for a single step, but only three in a long-running
simulation because the first force evaluation of a step occurs at the
same position as the last force evaluation of the previous step.  The
algorithm exactly conserves phase-space volume and momentum, but is
only fourth-order symplectic (in $2n$-dimensional phase-spaces with $n
> 1$).  As stated above, in practice we find that the energy error
from this algorithm in $N$-body simulations is significantly worse
than that from the following algorithm.  We compare the energy error
behavior of the algorithms in Figure \ref{OmelyanVsUs}.

\subsection{Prediction Algorithm}
\label{OurIteration}

As we shall see in Section \ref{IndividualTimeSteps}, it is not, in
general, possible to iterate equation \eqref{implicit} in the presence
of individual timesteps.  Iterating equation \eqref{implicit} for a
particle corresponds to re-running the evolution of that particle over
the first half of its timestep.  This may require re-running many
entire steps of particles with smaller timesteps than the given
particle, which, in turn, may require re-running yet more steps of
particles with even smaller timesteps.  The explosion of work is
exponential in the number of distinct timesteps assigned to particles
in the simulation.

Given that we cannot iterate equation \eqref{implicit}, it makes sense
to try to use all the information at hand to predict the solution $q'$
as well as possible at the beginning of a particle's timestep.  The
initial guess
\begin{equation}
\label{threePointPredictor}
q'_{(0)} = q_1 + \frac{h}{2} \frac{p_1}{m} + \frac{1}{2} \left(
\frac{h}{2} \right)^2 F(t) + \frac{1}{6} \left( \frac{h}{2} \right)^3
DF(t) + \frac{1}{12} \left( \frac{h}{2} \right)^4 D^2 F(t),
\end{equation}
where $t$ is the time at which the particle is at position $q_1$, and
$F(t)$ is the force on the particle as a function of time, solves
equation \eqref{implicit} with an error term of order $h^5$.  (Note
that the final term in equation \eqref{threePointPredictor} is twice
what would be expected in a power series for $q(t+h/2)$.)  We can use
the force evaluations from the previous step, at times $t-h$, $t-h/2$
and $t$, to approximate $F(t)$, $DF(t)$ and $D^2F(t)$ to sufficient
accuracy to compute equation \eqref{threePointPredictor} with
no new force evaluations.

\begin{figure}
  \begin{center}
    \plotone{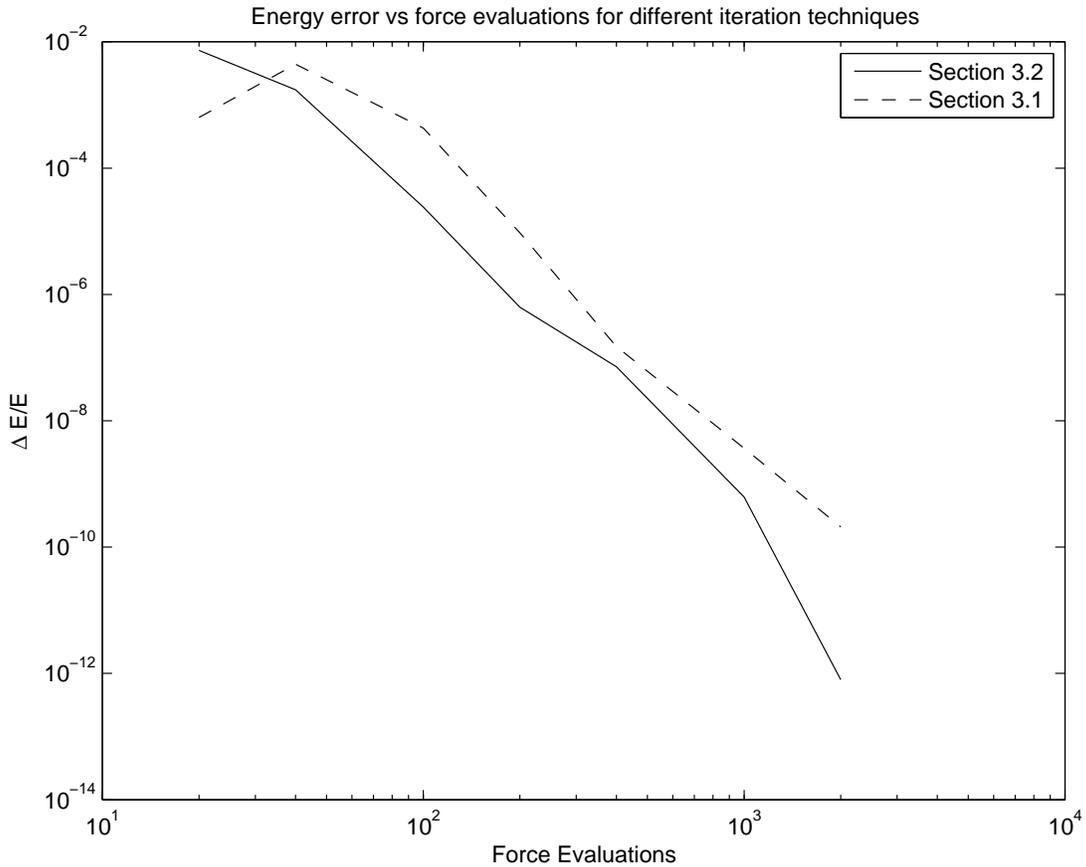}
  \end{center}
  \caption{\label{OmelyanVsUs} Energy error versus number of force
    evaluations for the algorithms described in Sections
    \ref{OmelyanIteration} and \ref{OurIteration} in a simulation of a
    100-body Plummer model initial condition with constant, shared
    timesteps over a total time interval $T = 1.0$.  As described in
    the text, the algorithm in Section \ref{OurIteration} outperforms
    the compositional algorithm in Section \ref{OmelyanIteration} for
    equivalent numbers of potential evaluations, though both
    algorithms exhibit the same asymptotic scaling.}
\end{figure}

The integration algorithm
\begin{eqnarray}
q_2 & = & q_1 + h \frac{p_1}{m} - \frac{1}{2} h^2 \left[
\frac{DV\left(q_1\right)}{3m} + \frac{2 DV\left( q'_{(0)} \right)}{3m}
\right] \\
p_2 & = & p_1 - h \left[ \frac{DV\left(q_1\right)}{6} + \frac{2
  DV\left( q'_{(0)} \right)}{3} + \frac{DV\left( q_2 \right)}{6} \right],
\end{eqnarray}
which is equations \eqref{3ptEquations} with $q'$ replaced by the
predicted $q'_{(0)}$ from equation \eqref{threePointPredictor}, is
fifth-order accurate in symplecticity and angular momentum
conservation, exactly conserves linear momentum, and requires only two
potential evaluations per step in a long-running simulation.  It
outperforms the algorithm in Section \ref{OmelyanIteration} in energy
error for equivalent force evaluations by one to two orders of
magnitude, as evidenced in Figure \ref{OmelyanVsUs}.  This is the
algorithm we shall discuss for the remainder of the paper.

\section{Adaptive Timesteps}
\label{AdaptiveTimesteps}

In $N$-body simulations it is essential to be able to adapt the
timestep taken by an integrator to the local conditions of the
system.  Formally, we can model such an adaptive step by treating
the parameter $h$ in the discrete Lagrangian describing an
integrator as a function of the positions through the step:
\begin{equation}
H \left( h, q_1, q'_1, \ldots, q^{(n)}_1, q_2 \right) \to H \left(
h\left( q_1, q'_1, \ldots, q_1^{(n)}, q_2 \right), q_1, q'_1,
\ldots, q^{(n)}_1, q_2 \right).
\end{equation}
Because the sequence of positions $q_1, q'_1, \ldots, q_1^{(n)},
q_2$ encodes all available information about the trajectory,
essentially \emph{any} technique for choosing a timestep can be
recast in this fashion.  We do not change the integrator equations
\begin{eqnarray}
\label{adpP1}
\partial_1 H \left( h\left( q_1, q'_1, \ldots, q^{(n)}, q_2 \right), q_1, q'_1, \ldots, q^{(n)}_1, q_2 \right) & = & -p_1 \\
\label{adpMiddlePoints}
\partial_i H \left( h\left( q_1, q'_1, \ldots, q^{(n)}, q_2 \right), q_1, q'_1, \ldots, q^{(n)}_1, q_2 \right) & = & 0, \qquad i = 2,3,\ldots,n+1 \\
\label{apdP2}\partial_{n+2} H \left( h\left( q_1, q'_1, \ldots, q^{(n)}, q_2 \right), q_1, q'_1, \ldots, q^{(n)}_1, q_2 \right) & = & p_2.
\end{eqnarray}

Assuming that the timestep function $h\left( q_1, q'_1, \ldots,
q_1^{(n)}, q_2 \right)$ is invariant under the same coordinate
variations which leave $H$ invariant (this will be the case if both
$H$ and $h$ inherit the same symmetries from the continuous
Lagrangian), then the proof of momentum conservation in equation
\eqref{momentumConservation} still holds.  All terms in $\delta H$
proportional to $\partial_0 H \left( h\left( q_1, q'_1, \ldots,
q_1^{(n)}, q_2 \right), q_1, q'_1, \ldots, q^{(n)}_1, q_2 \right)$
vanish because $\delta h$ also vanishes.  Adaptive stepping poses no
threat to momentum conservation, provided that the steps are chosen
in a way which respects the symmetries in the continuous mechanical
problem.

Unfortunately, adaptive stepping does pose a threat to
symplecticity.  With a timestep function $h\left( q_1, q'_1, \ldots,
q_1^{(n)}, q_2 \right)$, the expression for $d \mathcal{S}$
(see eq.\ \eqref{dS}) gets additional terms:
\begin{eqnarray}
\label{additionalAdaptiveTerms} 
\lefteqn{\sum_{i = 1}^{M - 1} \partial_0 H\left( h\left( q_i, q'_i, \ldots,
q_i^{(n)}, q_{i+1}\right), q_i, q'_i, \ldots, q_i^{(n)}, q_{i+1}
\right)} \nonumber \\ & \mbox{} \times \left[ \partial_0 h\left( q_i,
  q'_i, \ldots, q_i^{(n)}, q_{i+1}\right) dq_i + \sum_{j=1}^{n}
  \partial_j h\left( q_i, q'_i, \ldots, q_i^{(n)}, q_{i+1}\right)
  dq^{(j)}_i \right. & \nonumber \\ & \left. \mbox{} + \partial_{n+1}
  h\left( q_i, q'_i, \ldots, q_i^{(n)}, q_{i+1}\right)
  dq_{i+1}\right].
\end{eqnarray}
These extra terms prevent us from writing $d^2\mathcal{S} = 0$ as
the difference of two forms, one of which is the pushforward of the
other, as we did in equation \eqref{symplecticity}.  With general
adaptive timesteps, there is no two-form on the state space which is
conserved over a fixed number of steps.

This can be understood intuitively in the following way.
The conservation of the Poincar\'{e} integral
invariant,
\begin{equation}
\mathcal{I} \equiv \sum_i dq^i \wedge dp_i,
\end{equation}
with the sum taken over coordinate dimensions, along any trajectory
implies symplecticity (recall that eq.\ \eqref{dqWedgedpConservation}
shows that conservation of the two-form $\partial_1 \partial_{n+2}
H\left( h, q_1, q'_1, \ldots, q_1^{(n)}, q_2 \right) dq_1 \wedge dq_2$
on state space implies conservation of $\mathcal{I}$ on phase
space). Note that $dq^i$ and $dp_i$ are exterior derivatives and that
$\mathcal{I}$ is a two-form.  The Poincar\'{e} integral invariant
measures the sum of the areas of a tube of trajectories
infinitesimally near a reference trajectory projected onto the
sub-phase-spaces $\left(q^i, p_i\right)$.  But an integrator with an
adaptive timestep does not advance all trajectories in the tube with
the same $h$.  Even if we had an adaptive timestep integrator which
implemented the exact, continuous evolution of the system it still
would not conserve $\mathcal{I}$ over a single step---stopping the
evolution of the different trajectories in the tube at different times
spoils the symplecticity of the continuous evolution.  For an adaptive
timestep integrator symplecticity after a fixed number of steps
\emph{is the wrong condition to consider}.  Rather, we should ask
whether the integrator conserves the symplectic form over a fixed
total time of evolution.  In general, any well-behaved\footnote{In
  this case, ``well-behaved'' implies an integrator whose maps
  converge uniformly at order $h^r$ to the exact evolution map and
  whose derivatives converge uniformly as well.}  integrator
(including a variational integrator with general adaptive stepsizes)
which has trajectory error of order $h^r$ conserves the symplectic
form at least to order $h^r$ in this sense.

For variational integrators which choose timesteps using the popular
block-power-of-two scheme we can do better: these integrators conserve
the symplectic form almost everywhere in phase space.  In the
block-power-of-two scheme, a function $h_{\rm{max}}\left( q_1,
q'_1, \ldots, q_1^{(n)}, q_2 \right)$ limits the maximum timestep; the
actual timestep taken is rounded down from $h_{\rm{max}}$ to
the nearest number of the form $T/2^n$, with $n$ an integer, and $T$
some total evolution interval.  If the function $h_{\rm{max}}$
is continuous, about every point in state space for which
$h_{\rm{max}}\left( q_1, q'_1, \ldots, q_1^{(n)}, q_2 \right)
\neq T/2^n$ for some $n$ there is an open neighborhood of points which
round down to the same timestep.  Thus, the actual timestep function
$h\left(q_1, q'_1, \ldots, q_1^{(n)}, q_2 \right)$ is
piecewise-constant on state space, and the derivatives in equation
\eqref{additionalAdaptiveTerms} vanish almost everywhere on state
space for each step.  A variational integrator with adaptive
block-power-of-two timesteps is therefore symplectic almost everywhere
on state space because it is a composition of symplectic steps.
(``Almost everywhere'' should be taken in the mathematical sense of
``except on a set of measure zero.'')  In Section
\ref{SimulationResults} we present numerical evidence of this
phenomenon.  Incidentally, block-power-of-two timesteps have
advantages for parallelization which have led other authors to adapt
time-symmetric methods to block-power-of-two timesteps
\citep{Makino1996}.  Variational integrators fit naturally into a
block-power-of-two scheme which, in addition to its ease of
parallelization, preserves symplecticity in the presence of adaptive
timesteps.

\section{Individual Time Steps}
\label{IndividualTimeSteps} 

Astrophysical simulations of realistic systems often require
individual time steps for each body or for subgroups of bodies.  A
variation of the three-point GGL integrator described in Section
\ref{OurIteration} allows for individual steps.  This appears to be
the first individual timestep Runge-Kutta-Nystr\"{o}m integrator to
appear in the literature.

Our generalization to individual timesteps will exploit the way in
which we derived the 3-point GGL integrator.  In the derivation, we
assumed a quadratic-in-time trajectory for the particle which
interpolates between the positions $q_1$ at $t_1$, $q'$ at $t_1 +
h/2$, and $q_2$ at $t_1 + h$.  We will write 
\begin{equation}
q(t) = \phi\left( t; q_1, q', q_2, t_1, h \right)
\end{equation}
to represent this interpolated trajectory in what follows.  We now
repeat the derivation which led to equations \eqref{3ptEquations},
allowing each body to have a different timestep.

Write the $N$-body Lagrangian as a sum over particle indices $i = 1
\ldots N$:
\begin{equation}
\label{itLagrangian} L  = \sum_{i = 1}^N \left[ \frac{m^i}{2}
\left(Dq^i(t) \right)^2 - \sum_{j = i+1}^N m^i m^j V\left(q^i(t) -
q^j(t) \right) \right] \equiv \sum_{i=1}^N \mathcal{T}^i -
\mathcal{V}^i
\end{equation}
where $V(q) = -1/|q|$ (recall we suppress vector indices).  Assume for
the moment that the timesteps of each body, $h^i$, are fixed in time
(Section \ref{Implementation} discusses how to integrate adaptive
steps into this algorithm), and that particles are sorted so that $h^i
\leq h^j$ when $i < j$.  Note that $\mathcal{T}^i$ depends only on the
trajectory of body $i$, while $\mathcal{V}^i$ depends on the
trajectories of bodies $j$ with $j \geq i$.

\begin{figure}
\begin{center}
\plotone{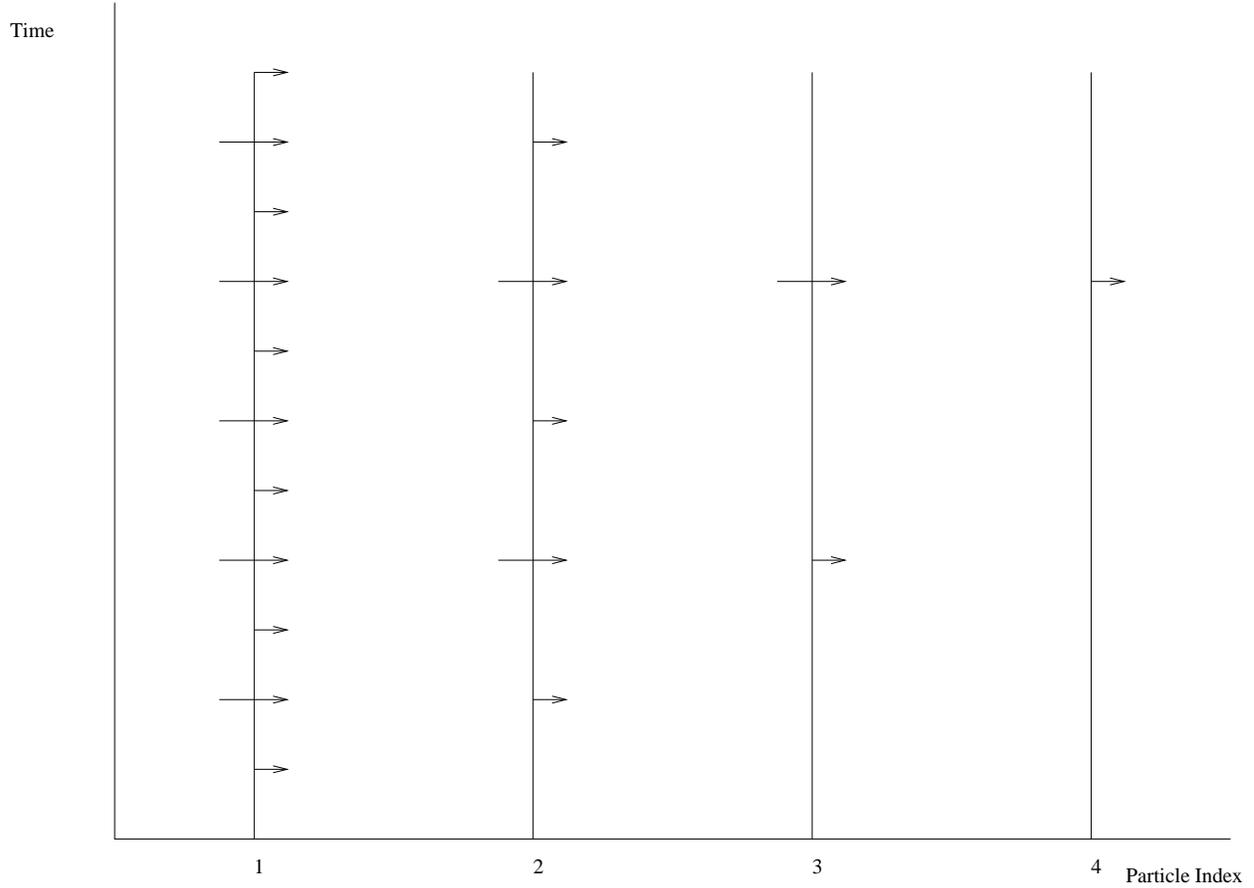}
\end{center}
\caption{\label{potentialEvaluations} Time-sequence of potential
  evaluations in the integral approximation to the action, equation
  \eqref{itActionQuadrature}, assuming, for graphical simplicity, that
  $h^j = 2^{j-1} h^1$.  Time runs vertically and particle index runs
  horizontally.  A right-facing arrow crossing a particle timeline
  represents evaluation of potentials involving that particle and all
  particles of greater index.  Long arrows represent evaluations of
  potentials at the beginning and end of particle timesteps
  (associated with coefficients $h/6$); short arrows represent
  evaluations of potentials at particle half-timesteps (with
  coefficient $2h/3$). }
\end{figure}

Now compute the discrete approximation to the action over a total time
interval $\Delta t$ using the three-point Gauss-Lobatto quadrature
rule to integrate each of the $\mathcal{T}^i - \mathcal{V}^i$:
\begin{equation}
\label{itActionQuadrature}
S \approx \sum_{i=1}^N \sum_{k = 0}^{n^i - 1} h^i \left(
\left. \frac{\mathcal{T}^i - \mathcal{V}^i}{6} \right|_{t = kh^i} +
\left. \frac{2 \left(\mathcal{T}^i - \mathcal{V}^i\right)}{3}
\right|_{t = \frac{2k+1}{2} h^i} + \left. \frac{\mathcal{T}^i -
  \mathcal{V}^i}{6} \right|_{t = (k+1)h^i} \right),
\end{equation}
where $\Delta t = n^i h^i$ for each $h^i$.  Recall that
$\mathcal{V}^i$ contains contributions from potentials $V\left( q^i(t)
- q^j(t)\right)$ with $j > i$; this sequence of potential evaluations
is illustrated in Figure \ref{potentialEvaluations}.

\begin{figure}
\begin{center}
\plotone{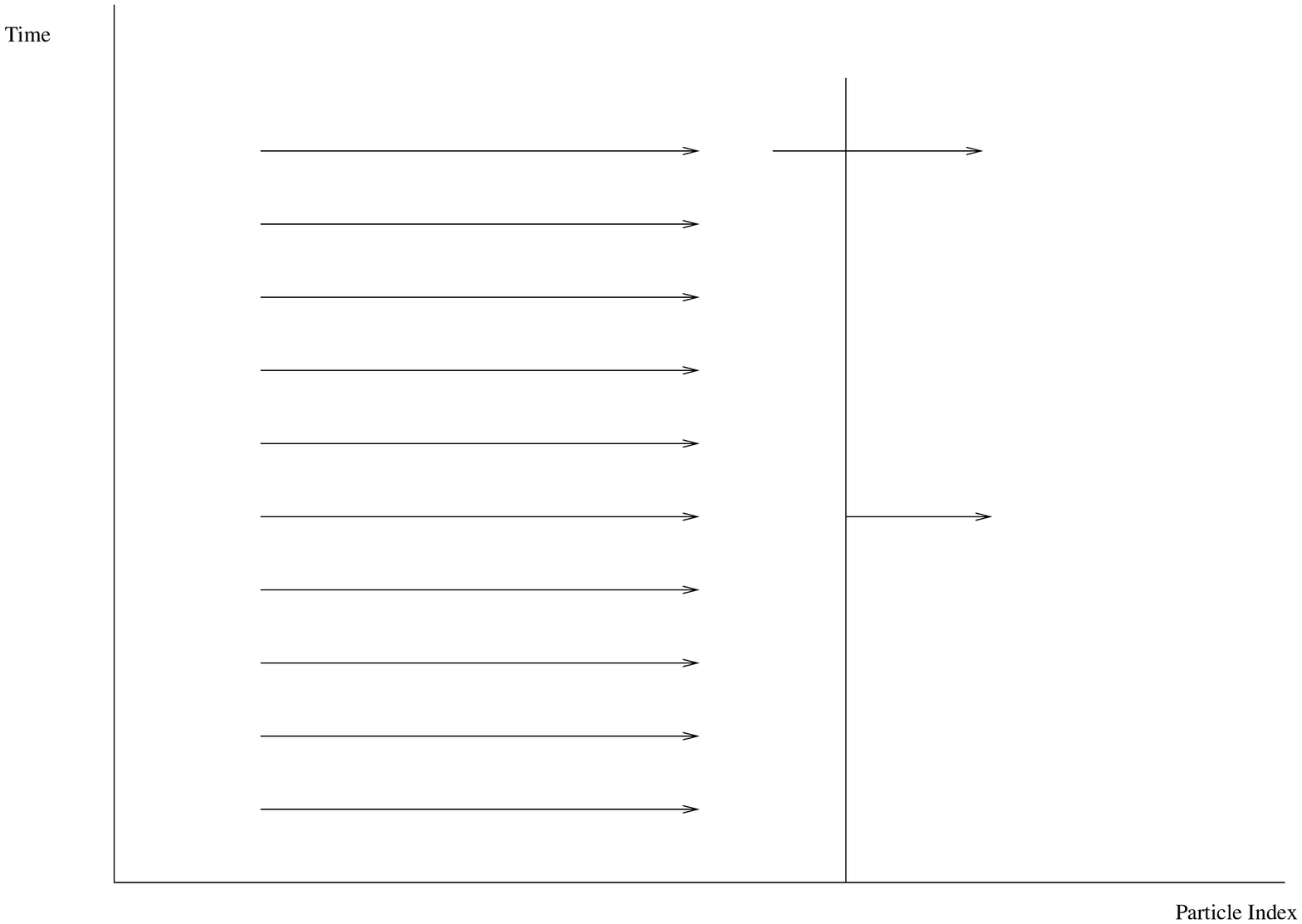}
\end{center}
\caption{\label{particleH} Graphical representation of the potential
  evaluations involving three particular interpolation points,
  $q^i_k$, $q'^i_k$ and $q^i_{k+1}$ for particle $i$.  Time flows
  vertically, and particle index runs horizontally.  Potential
  evaluations in $\mathcal{V}^i$ go to the right; each of these
  evaluations involves exactly one of the positions $q^i_k$, $q'^i_k$
  and $q^i_{k+1}$.  Potential evaluations in $\mathcal{V}^j$, $j < i$,
  come from the left, and generally involve the interpolated position
  of particle $i$.}
\end{figure}

Discretize the trajectories of each particle using a quadratic
interpolation between the integration points:
\begin{equation}
q^i(t) = \phi\left(t; q^i_k, q'^i_k, q^i_{k+1}, k h^i, h^i \right),
\quad k h^i \leq t \leq (k+1)h^i.
\end{equation}
A given interpolation point, $q^i_k$, $q'^i_k$ or $q^i_{k+1}$, will
appear in $\mathcal{T}^i$ and $\mathcal{V}^i$ on step $k$ of index
$i$, and $\mathcal{V}^j$, $j < i$, on all steps $l$ such that $l h^j
\leq (k+1) h^i$ and $(l+1) h^j \geq k h^i$.  In general, $q^i_k$,
$q'^i_k$ or $q^i_{k+1}$ appear together in $\mathcal{V}^j$ as part of
an interpolation $\phi$, while each appears once alone in the three
$\mathcal{V}^i$ terms of timestep $k$.  Figure \ref{particleH}
illustrates graphically the time sequence of all potential evaluations
involving three particular interpolation points, $q^i_k$, $q'^i_k$ and
$q^i_{k+1}$.

Extremizing the action with respect to $q^i_k$, $q'^i_k$ and
$q^i_{k+1}$ as in Section \ref{VariationalIntegrators}, and using
equation \ref{threePointPredictor} to predict $q'^i_k$ as 
\begin{equation}
q'^i_k = q^i_k + \frac{h^i}{2} \frac{p^i_k}{m} + \frac{1}{2} \left(
\frac{h^i}{2} \right)^2 F\left( k h^i \right) + \frac{1}{6} \left(
\frac{h^i}{2} \right)^3 DF\left( k h^i \right) + \frac{1}{12} \left(
\frac{h^i}{2} \right)^3 D^2F\left( k h^i \right),
\end{equation}
gives 
\begin{mathletters}
\label{itRKNEquations}
\begin{eqnarray}
q^i_{k+1} & = & q^i_k + h^i \frac{p^i_k}{m} - \frac{1}{2}
\left(h^i\right)^2 \left[ \frac{1}{3m} \left( \frac{\partial
  \mathcal{V}^i}{\partial q^i_k} + \sum_{j < i} \frac{\partial
  \mathcal{V}^j}{\partial q^i_k} \right) + \frac{2}{3m} \left(
\frac{\partial \mathcal{V}^i}{\partial q'^i_k} + \sum_{j < i}
\frac{\partial \mathcal{V}^j}{\partial q'^i_k} \right) \right] \\
p^i_{k+1} & = & p^i_k - h^i \left[ \frac{1}{6m} \left( \frac{\partial
  \mathcal{V}^i}{\partial q^i_k} + \sum_{j < i} \frac{\partial
  \mathcal{V}^j}{\partial q^i_k} \right) + \frac{2}{3m} \left(
\frac{\partial \mathcal{V}^i}{\partial q'^i_k} + \sum_{j < i}
\frac{\partial \mathcal{V}^j}{\partial q'^i_k} \right) \right. 
\nonumber \\
& & \left. + \frac{1}{6m} \left( \frac{\partial
  \mathcal{V}^i}{\partial q^i_{k+1}} + \sum_{j < i} \frac{\partial
  \mathcal{V}^j}{\partial q^i_{k+1}} \right) \right].
\end{eqnarray}
\end{mathletters}
We can evaluate the force terms in the above equations in the same
time sequence as the potential evaluations in equation
\eqref{itActionQuadrature}.  First, assume we have predicted
$q^i_{k+1}$ using $F\left(kh^i\right)$, $DF\left(kh^i\right)$ and
$D^2F\left(kh^i\right)$ in addition to predicting $q'^i_k$.  As we
advance the system forward in time (imagine a horizontal line rising
upward in Figure \ref{potentialEvaluations}), we encounter potential
evaluations between various pairs of particles $i$ and $j$.  When we
encounter a potential evaluation, we must compute the corresponding
contribution to the $\partial \mathcal{V}/\partial q$ terms in
equations \eqref{itRKNEquations}.  We use the chain rule: each
$\partial \mathcal{V}/\partial q$ term is the product of a force
evaluated at interpolated positions and the interpolation coefficient
of the corresponding coordinate in the position interpolation.  We
accumulate all such contributions over the course of each timestep.

When we reach the end of a particular particle's step, we use
equations \eqref{itRKNEquations} to advance the position and momentum
of that particle, and then predict the new $q'^i_{k+1}$ and
$q^i_{k+2}$ using the stored force data.  Each particle must store its
mass, its timestep, $h^i$, three positions for the timestep, $q^i_k$,
$q'^i_k$ and $q^i_{k+1}$, the initial momentum $p^i_k$, and three
force accumulators, $\partial \mathcal{V}/\partial q^i_k$, $\partial
\mathcal{V}/\partial q'^i_k$ and $\partial \mathcal{V}/\partial
q^i_{k+1}$, for a total storage of 23 floating-point numbers in three
dimensions (in double precision this is 184 bytes per particle).

\section{Adaptive Stepping and Individual Block Power of Two Timesteps}
\label{Implementation} In this section we combine the
block-power-of-two adaptive timesteps described in Section
\ref{AdaptiveTimesteps} with the individual timestep algorithm
described in Section \ref{IndividualTimeSteps}.  In order that
equation \eqref{itActionQuadrature} approximate the action for the
$N$-body system, we require $h^i \leq h^j$ for $i < j$.  Thus,
changing the timestep for a body potentially requires re-ordering the
sum in equation \eqref{itLagrangian}.  Bodies $i$ and $j$ can only
change relative position in the sum when \emph{both} have finished a
step, or there will be some potentials whose evaluations in the
discrete Lagrangian do not correspond to a proper Gauss-Lobatto
quadrature rule.  Using block-power-of-two timesteps,
\begin{equation}
\label{po2}
h^i = \frac{\Delta t}{2^{p^i}}, \quad p^i = 1 {\rm\ or\ } 2
{\rm\ or\ } 3 \ldots,
\end{equation}
for some total integration interval $\Delta t$.  With
block-power-of-two timesteps, all bodies completing a timestep at a
given time are contiguous in the sum from indices $1$ to some
$i^{\rm{max}}$.  We can re-compute the maximum timestep
allowable for each of these bodies and sort them in the sum
according to their maximum timestep.  Once the maximum timesteps are
calculated, we must ensure that the actual timesteps taken satisfy
\begin{equation}
h^i \leq \min \left( h^i_{\rm{max}}, h^{i^{\rm{max}}
+ 1} \right), \quad i = 1, 2, \ldots, i^{\rm{max}},
\end{equation}
subject to the power-of-two restrictions in equation  \eqref{po2}.
This procedure preserves the invariant that $h^i < h^j$ for $i < j$
and ensures $h^i < h^i_{\rm{max}}$.

\citet{Makino1991} recommends choosing a timestep which reflects the
error of the predictor relative to the final solution.  We choose
\begin{equation}
\label{timestep} h^i_{\rm{max}} = h^i \left( \frac{\eta
\left(h^i\right)^2 F}{m \left| q^i - q^i_{\rm{pred}}
\right|} \right)^{1/5}
\end{equation}
where $\eta$ is an accuracy parameter of the simulation, and
$q^i_{\rm{pred}}$ is the predicted position of the particle
using equation \eqref{threePointPredictor}, while $q^i$ is the
position as determined by the integrator at the end of the prior
timestep.  The exponent is $1/5$ because the predictor has an $O(h^5)$
error term.

The complete algorithm to advance body $i$ by $dt$, assuming an
array of bodies sorted by $h_{\rm{max}}$ and indexed from $0$
to $N-1$ is then:
\begin{enumerate}
\item If $dt \leq h^i_{\rm{max}}$ then
\begin{enumerate}
\item Predict the positions of body $i$ at $t + dt/2$ and $t + dt$
  using equation \eqref{threePointPredictor}, and a Taylor series for
  $q(t+dt)$, respectively.
\item Compute potential gradients between body $i$ and bodies $j$,
  with $j > i$, at times $t$, $t + dt/2$, and $t + dt$, distributing
  the forces across the $\partial \mathcal{V}/\partial q$ accumulators
  of body $j$ according to the interpolation coefficients for the
  position of body $j$ at these times.  Also accumulate the forces into
  the corresponding accumulator of body $i$.
\item Unless $i=0$, advance body $i-1$ by $dt$.  Upon completion of
  this step the forces on body $i$ from all bodies with indices below
  $i$ will be stored in $i$'s accumulators.
\item Update the position of body $i$ using equations
  \eqref{itRKNEquations}.  Calculate a new $h^i_{\rm{max}}$.
\end{enumerate}
\item Otherwise
\begin{enumerate}
\item Advance body $i$ by $dt/2$.
\item Sort bodies $0$ to $i$ according to their maximum timesteps.
\item Advance the body at index $i$ (which may not be the same body after the sort) by $dt/2$.
\end{enumerate}
\end{enumerate}
The algorithm begins by attempting to advance body $N-1$ by the
entire time interval, $\Delta t$.  The order of computations and
recursion above ensures that the timesteps are block-power-of-two,
that the total integration terminates after advancing the bodies by
exactly $\Delta t$, and maintains the invariant that $h^i \leq h^j$
when $i < j$.

\section{Numerical Experiments}
\label{SimulationResults}

In this section we report on numerical experiments involving two
different GGL variational integrators.  The first integrator solves
for one-dimensional Keplerian orbits using the evolution mapping
defined by equations \eqref{3ptEquations} and the standard Kepler
Lagrangian; the second is a general $N$-body integrator using the
adaptive-stepsize, individual timestep algorithm described above.
The source code for both integrators is available upon request.

\subsection{One-Dimensional Simulation}
\label{oneDSimulation} 
\begin{figure}
\begin{center}
\plotone{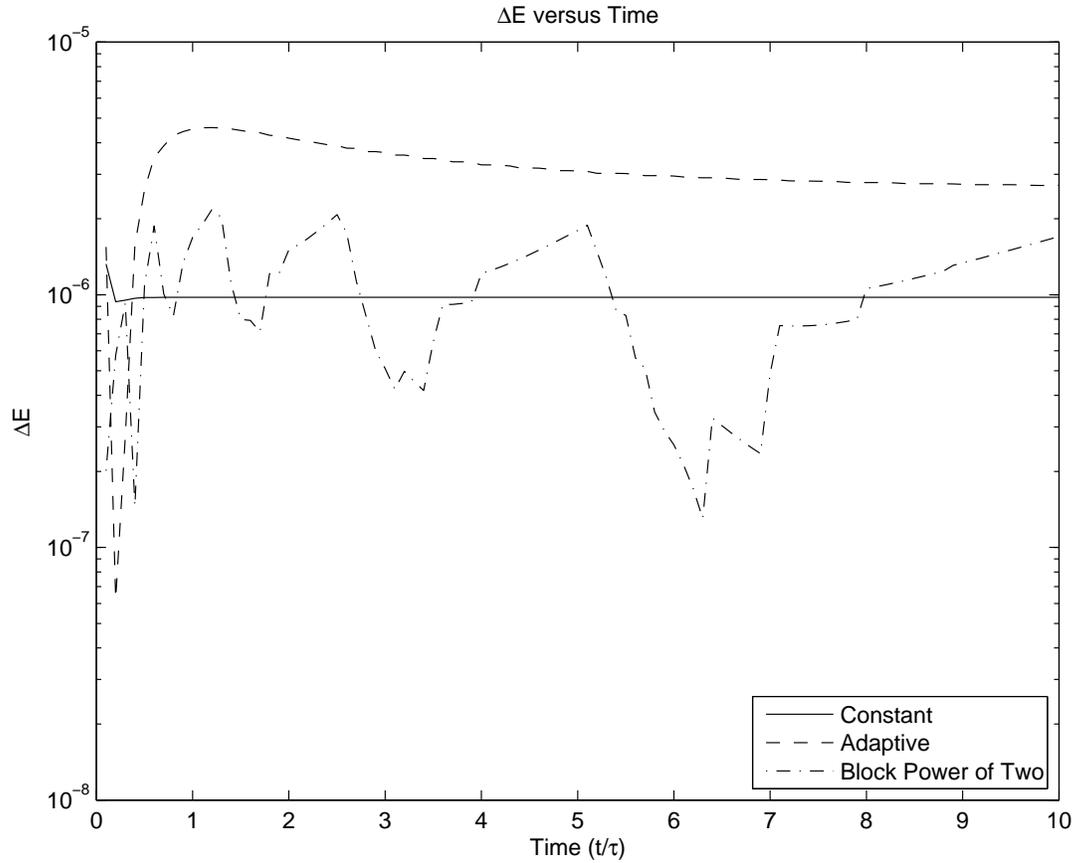}
\end{center}
\caption{\label{OneDEnergyError} Relative energy error versus total
  evolution interval for simulations of the Kepler orbit described in
  the text.  The orbital parameters are $a = 2/45$, $e = 9/10$, $E =
  -1/4$.  The total evolution interval for each simulation varies from
  $0$ to $10\tau$ in steps of $\tau/10$.  All three integration
  algorithms have comparable energy error.}
\end{figure}

\begin{figure}
\begin{center}
\plotone{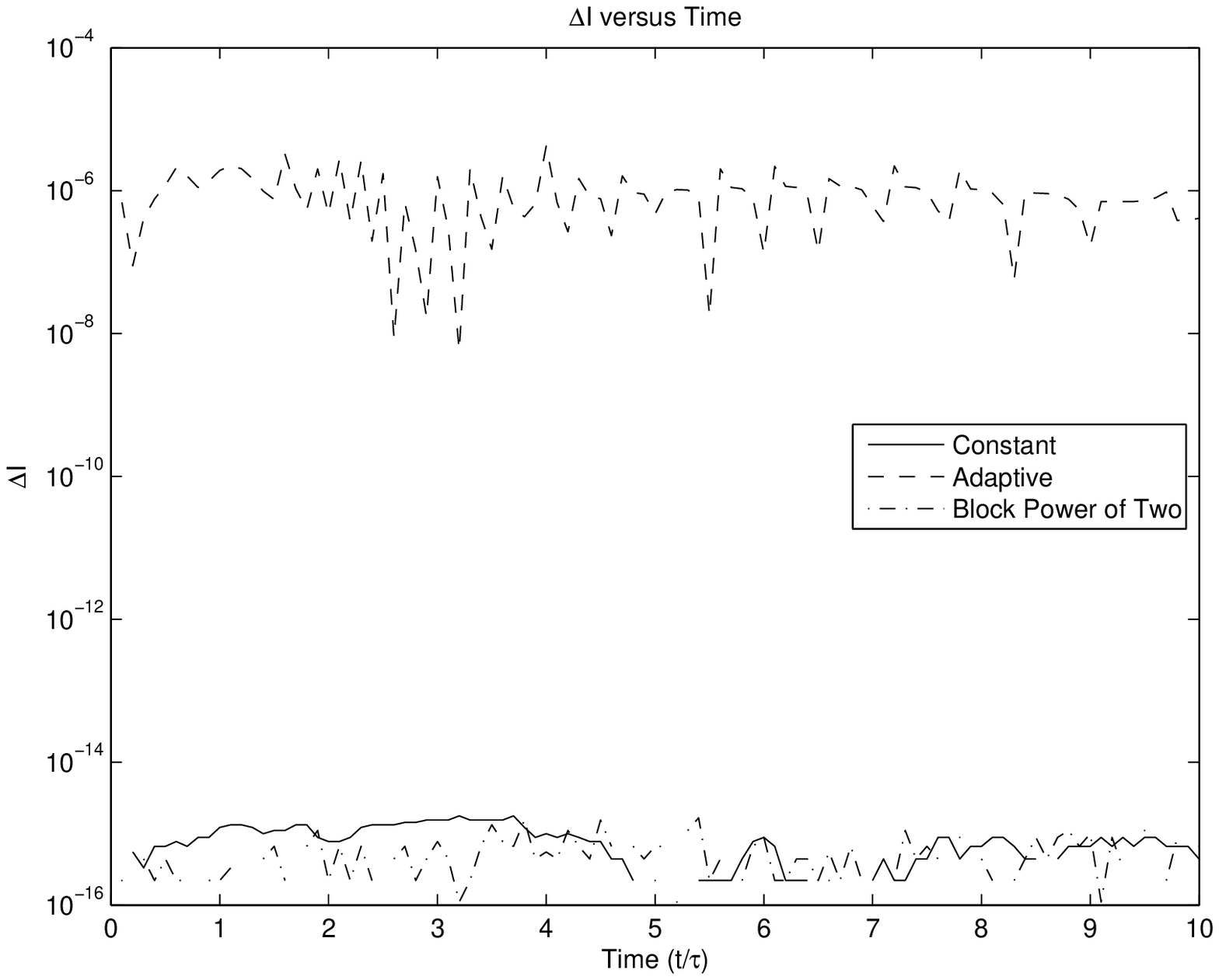}
\end{center}
\caption{\label{OneDSymplecticityError} Error in the Poincar\'{e}
  integral invariant versus total evolution interval for the
  simulations of the Kepler orbit shown in Figure
  \ref{OneDEnergyError}.  The constant timestep (solid line) and
  block-power-of-two timestep (dot-dashed line) algorithms conserve
  the Poincar\'{e} integral invariant exactly while the adaptive
  timestep algorithm (dashed line) doesn't, even though the energy
  error behaviors of the algorithms are comparable.}
\end{figure}

To illustrate the energy conservation and
symplecticity performance of adaptive and non-adaptive variational
integrators, we performed several one-dimensional simulations of the
Kepler problem,
\begin{equation}
L(r,v) = \frac{m v^2}{2} - \frac{l^2}{2 m r^2} + \frac{k}{r}.
\end{equation}
The particular parameters we chose were $m = k = 2/9$, and $l =
2\sqrt{19}/135$, with initial conditions $r_0 = 2/45$, $v_0=p_0=
0.0$.  With these parameters and initial conditions the total energy
of the system is $E = -1/4$, the eccentricity is $e = 9/10$, and the
orbital period is $\tau = 2\pi \left(2/45\right)^{3/2} \approx
0.0588716$.  We evolved the system for various times, from $\tau/10$
to $10 \tau$ in increments of $\tau/10$.  The plots shown in Figures
\ref{OneDEnergyError} and \ref{OneDSymplecticityError} use values
computed at the end of a simulation over the appropriate total time
interval, not snapshots of the corresponding values in an ongoing
simulation; the distinction is important because of what it implies
about the choices of adaptive timesteps---see Section
\ref{AdaptiveTimesteps}.  We used three different integration
algorithms based on equations \eqref{3ptEquations}: a constant
timestep integrator, an adaptive timestep integrator where $h$ is
chosen at the beginning of a step according to
\begin{equation}
h(q,p) = \eta \left( \frac{m q^3}{k} \right)^{1/2},
\end{equation}
where $q=r$ and we set $\eta = 0.05$, and a block-power-of-two
timestep integrator which uses the above for its
$h_{\rm{max}}$.  For the constant timestep integrator, we
chose the timestep $h\left( r_0, p_0 \right)$.  All algorithms iterate
equation \eqref{implicit} to convergence; we thus expect the
block-power-of-two and constant timestep algorithms to be exactly
symplectic.

Figure \ref{OneDEnergyError} displays the relative energy error
accumulated over many simulated orbits by the three algorithms.  It
is clear that the three algorithms accumulate comparable energy
error, with the adaptive timestep choice slightly worse than the
others.

Given an evolution mapping $F_t : (q,p) \mapsto (Q_t(q,p), P_t(q,p))$,
the pushforward of the Poincar\'{e} integral invariant $\mathcal{I}
\equiv dq \wedge dp$ along the trajectory $q(t), p(t) = F_t\left(q_0,
p_0 \right)$ is given by
\begin{equation}
\label{OneDSymplecticPushforward} \left(F^*_t
\mathcal{I}\right)\left(q_0, p_0\right) = \left[ \partial_0
Q_t\left( q_0, p_0 \right) \partial_1 P_t\left( q_0, p_0 \right) -
\partial_1 Q_t\left( q_0, p_0 \right) \partial_0 P_t \left( q_0, p_0
\right) \right] dq_0 \wedge dp_0.
\end{equation}
All three integration algorithms are evolution mappings.  An elegant
algorithm \citep{Sussman2006,Sussman2001} exists to compute
derivatives of arbitrary computations, such as our evolution mappings,
without the truncation error which would result from finite
differencing.  We used this algorithm to compute $\Delta I \equiv
\left[\left(F^*_t \mathcal{I}\right) - \mathcal{I}\right]\left(q_0,
p_0 \right)$ using equation \eqref{OneDSymplecticPushforward}.
$\Delta I$ is a two form, and in a two-dimensional phase space it can
be written $\Delta I = f(t) dq_0 \wedge dp_0$, where $f$ is a scalar.
The value of $f$ is plotted versus $t$ for the three algorithms in
Figure \ref{OneDSymplecticityError}.  It is clear that the ordinary
adaptive stepsize integrator does not conserve the symplectic form
while the block-power-of-two and constant stepsize integrators do, as
expected from Section \ref{AdaptiveTimesteps}.  As explained in
Section \ref{AdaptiveTimesteps}, though the energy error performance
of the three algorithms is comparable, only the constant-timestep and
block-power-of-two schemes are symplectic, and only the
block-power-of-two scheme is symplectic \emph{and} allows for adaptive
timesteps.

\subsection{Many-Body Simulations}
In this subsection we report on the application of the individual
and adaptive timestep integration algorithm described in this paper
to simulations of many-body systems.  We choose to use the so-called
``standard units'': units in which the total system mass $M = 1$, $G
= 1$, and the total energy $E = - 1/4$ \citep{Heggie1986}.  In
standard units, the virial radius of the system is 1.  Our initial
conditions are a randomly sampled Plummer model shifted into a
coordinate system where the center of mass is at the origin and the
total linear momentum is zero.

Because our code makes no special provision for close encounters, we
used a softened gravitational potential:\begin{equation} V(r) =
\frac{1}{\sqrt{r^2 + \epsilon^2}},
\end{equation}
with $\epsilon = 4/N$, where $N$ is the number of bodies in a
particular simulation.  This is a standard technique in codes which
do not carefully regularize the singular two-body potential
\citep{Aarseth2001}.

\subsubsection{Symplecticity Tests}
\label{SymplecticityTests}
An $N$-body system with $N$ bodies has a $6N$-dimensional phase
space.  Given an integration mapping, with Jacobian matrix $J$,
conservation of the Poincar\'{e} integral invariant by the mapping
implies that
\begin{equation}
\label{symplecticUnitConservation}
S = J^T S J,
\end{equation}
where $S$ is the ``symplectic unit'':
\begin{equation}
S = \left[
\begin{array}{cc}
0 & -1 \\
1 & 0
\end{array}
\right]
\end{equation}
where $1$ represents the identity matrix in $3N$-dimensions.  Using
a generalization of the algorithm for explicitly computing
derivatives of computational algorithms, we can compute the Jacobian
matrix for the evolution mapping defined by any integration scheme.

We denote by $dI$ the relative change in the sum of the absolute
values of the diagonal from lower left to upper right between $J^TSJ$
and $S$.  If equation \eqref{symplecticUnitConservation} holds, $dI =
0$.  We compare $dI$ from the individual and adaptive timestep
integrator described in this paper to $dI$ from a standard individual
and adaptive timestep Hermite integrator \citep{Makino1991}.  It is
difficult to precisely control the stepsizes chosen in an adaptive
timestep scheme, so we measure $dI$ as a function of the total number
of steps taken for the evolution, $n_{\rm{steps}}$.  An error
which scales as $h^r$ should scale as $n_{\rm{steps}}^{-r}$.

\begin{figure}
\begin{center}
\plotone{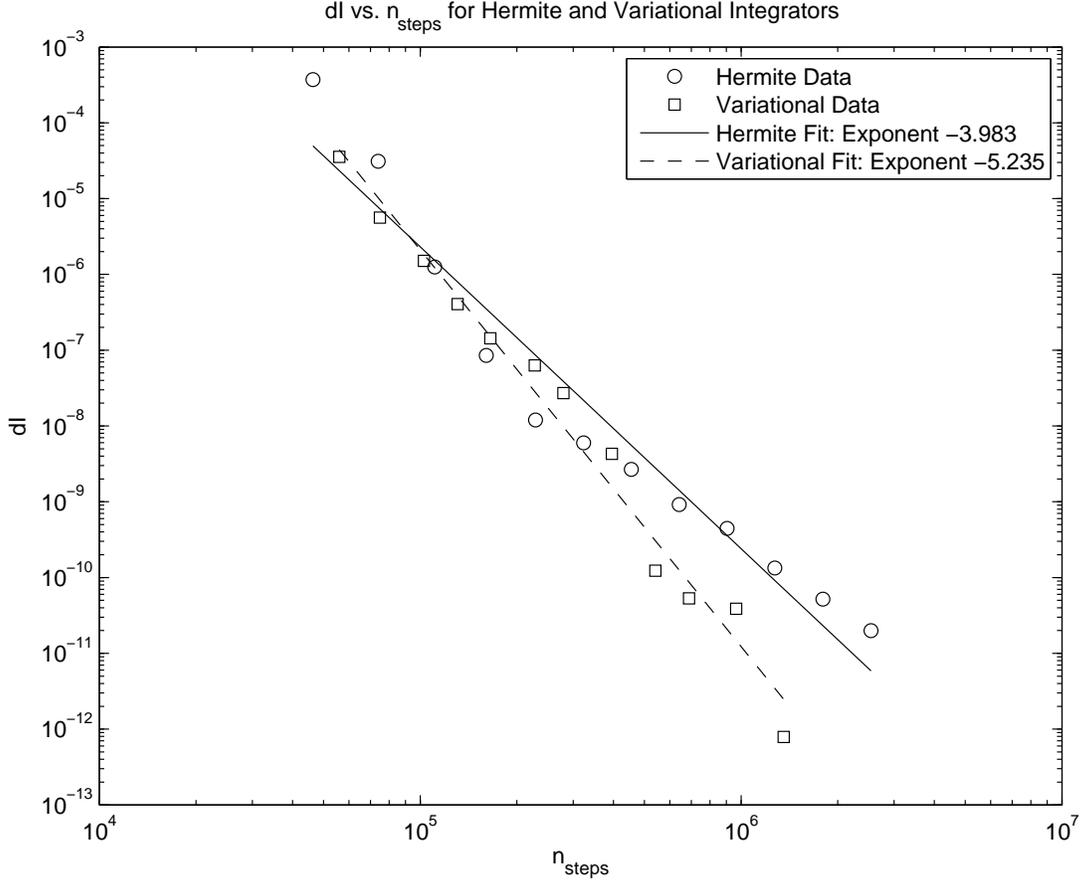}
\end{center}
\caption{\label{hermiteVariationalSymplecticError} Error in equation
  \eqref{symplecticUnitConservation} (non-conservation of the
  Poincar\'{e} integral invariant) in multiple integrations of $N=25$
  Plummer initial conditions by the standard Hermite integrator and
  the individual and adaptive timestepping integrator described in the
  text over a total time interval of $T = 1.0$ with various choices of
  (adaptive) timestep.  The data have been least-squares fit to
  illustrate the scaling with $n_{\rm steps}$: the Hermite data fit
  with $dI \propto n_{\rm steps}^{-3.983}$, and the variational data
  fit with $dI \propto n_{\rm steps}^{-5.235}$, as expected from the
  discussion in Section \ref{ThreePointIntegrator}.}
\end{figure}

Based on the discussion in Section \ref{ThreePointIntegrator}, we
expect that $dI$ from the variational integrator will be fifth order
(that is, it scales as $n_{\rm{steps}}^{-5}$), while $dI$ from the
Hermite integrator will be fourth order (exactly the same order as the
integrator's trajectory error).  Figure
\ref{hermiteVariationalSymplecticError} demonstrates that this is
exactly the case for simulations of a Plummer initial condition with
$N = 25$ bodies and varying numbers of steps over a total time
interval of $T = 1.0$ by both algorithms.  

Because they both depend on the accuracy of the solution to the
implicit equation \eqref{implicit}, we expect that symplecticity and
angular momentum conservation error would scale similarly in a
long-term simulation.  Results in the next section show that angular
momentum error in a long-running simulation is about one order of
magnitude better than the energy conservation error (Figure
\ref{ManyBodyDeltaEAndL}) using our algorithm, while the Hermite
algorithm produces angular momentum errors commensurate with its
energy conservation error.  Our algorithm appears to have the
advantage in angular momentum conservation and symplecticity in
large-$N$, long-time simulations.

\subsubsection{1000-body Cluster Simulation}
\label{ManyBodySimulation}
Here we compare a 1000-body cluster simulation run using our
variational integrator with another such simulation using the NBODY2h
code \citep{Aarseth2001}.  NBODY2h is not the state-of-the-art in
cluster simulations; it is an appropriate comparison to our code
because it uses the state-of-the-art Hermite integration algorithm but
uses softening instead of treating close encounters specially.  We
have run many such simulations---the one described here is typical.
Recall we use the standard units: the total system mass is $M = 1$, $G
= 1$, and the total energy $E = - 1/4$ \citep{Heggie1986}.  In
standard units, the virial radius of the system is 1.  Initial
conditions for both runs are a randomly sampled Plummer model shifted
into a coordinate system where the center of mass is at the origin and
the total linear momentum is zero.  The random sampling differs
between the two codes, so the initial conditions are not identical.

In the simulation reported here, we chose to keep energy
conservation to better than a part in $5\times 10^{-9}$ per unit
timestep, with a target of one part in $10^{-9}$.  If, at the end of
an advancement by $\Delta t = 1.0$, the relative energy error was
greater than $5\times10^{-9}$ we re-started the step with a smaller
$\eta$ parameter (cf.\ eq.\ \eqref{timestep}); at the end of every
unit time interval, we adjusted $\eta$ according to
\begin{equation}
\eta' = \eta \frac{10^{-9}}{\Delta E/E}.
\end{equation}
A similar procedure is used to adjust the timestep in NBODY2h.

\begin{figure}
\begin{center}
\plotone{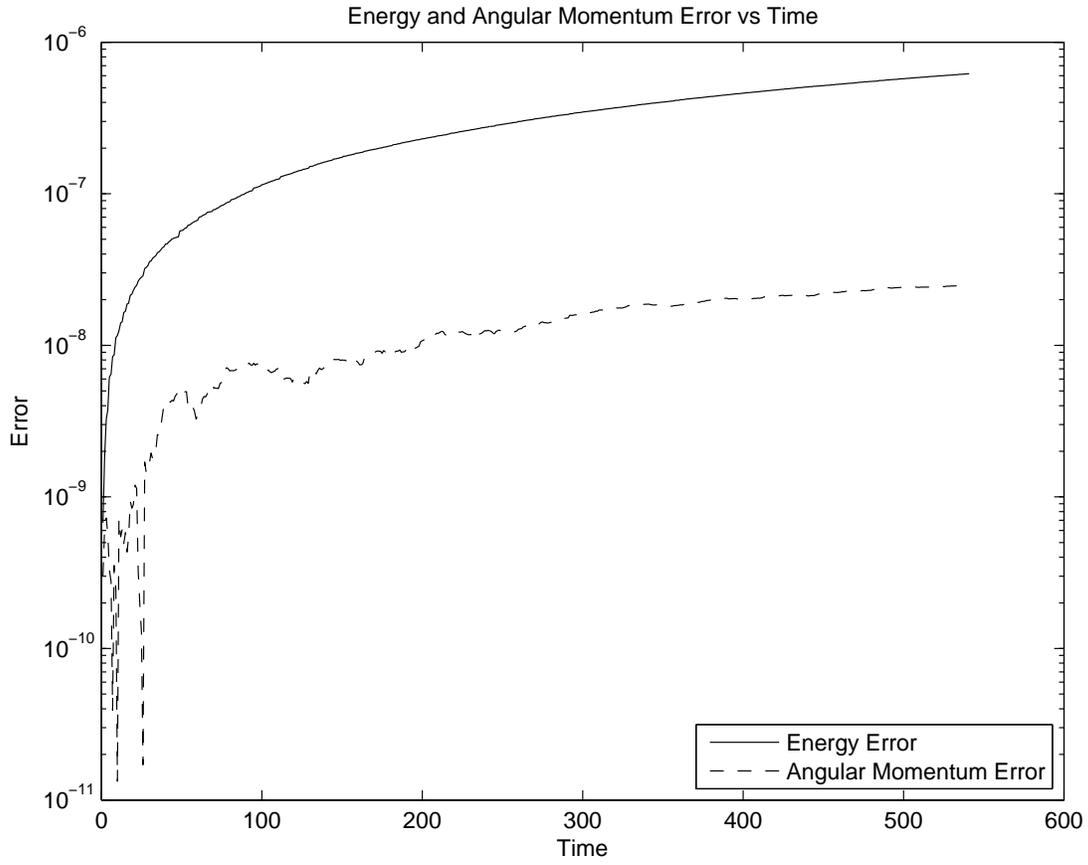}
\end{center}
\caption{\label{ManyBodyDeltaEAndL} Relative energy and angular
  momentum conservation error for the 1000-body simulation performed
  using the variational integrator.  As expected, the angular momentum
  conservation error is much better than the energy conservation
  error.}
\end{figure}

Figure \ref{ManyBodyDeltaEAndL} shows the time-evolution of the
relative energy error and relative angular momentum error in the
variational simulation.  We can see in Figure \ref{ManyBodyDeltaEAndL}
that the angular momentum conservation error is about an order of
magnitude smaller than the energy conservation error.  This is
consistent with the fact that the angular momentum error scales one
power of $h$ better than the energy error.

\begin{figure}
\begin{center}
\plotone{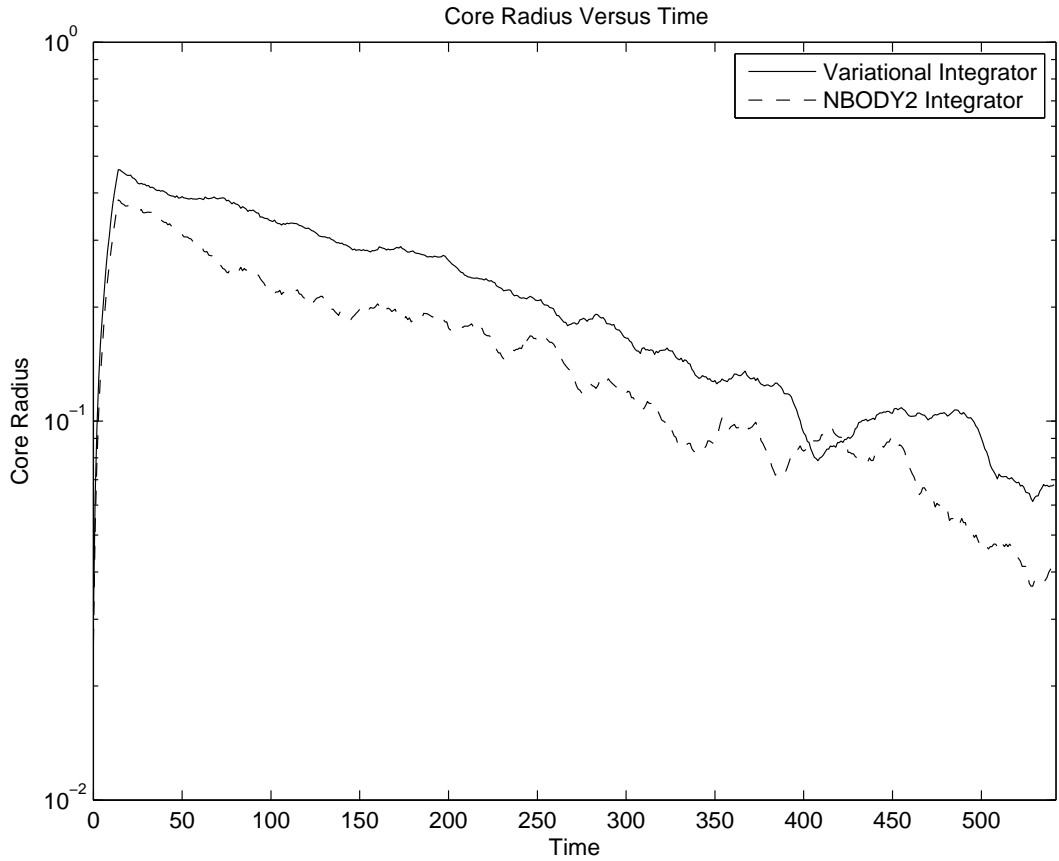}
\end{center}
\caption{\label{ManyBodyCoreRadius} Core radius versus time for the
  simulation described in the text using the integrator in this paper
  and an equivalent simulation using the NBODY2h code.  Boxcar
  averaging has been employed with $\Delta t = 15$ to reduce
  fluctuations in the curves.}
\end{figure}

Figure \ref{ManyBodyCoreRadius} compares the core radius (computed as
described in \citet{Casertano1985}) versus time for the variational
simulation with the core radius output by the NBODY2h simulation
(recall that the two simulations do not start with identical initial
conditions, but rather random samplings of identical phase-space
distributions).  The dynamical behavior of the core radius simulated
using our algorithm is qualitatively correct.  The NBODY2h simulation
has momentum (both linear and angular) conservation error which is
approximately commensurate with its energy conservation error---about
one order of magnitude worse than the variational algorithm's angular
momentum conservation error, and six orders of magnitude worse than
its linear momentum conservation error.  Its wall-clock time is
significantly better than the variational algorithm because our
algorithm does not implement the nearest-neighbor scheme of
\citet{Ahmad1973}, and therefore evaluates all 999 potentials
involving the body with the smallest timestep every step, all other
998 potentials involving the body with the second-smallest timestep
every step for that body, etc.

\section{Discussion}
\label{Conclusions}

We have presented a class of $N$-body integrators obtained by
discretizing the action as opposed to discretizing the equations of
motion.  Such integrators automatically conserve discrete momenta and
are symplectic. This paper focuses on the fourth-order, three-point
GGL integrator described in Section \ref{ThreePointIntegrator}, but
Section \ref{VariationalIntegrators} describes a general framework for
constructing such integrators. We have demonstrated, theoretically in
Section \ref{AdaptiveTimesteps} and experimentally in Section
\ref{oneDSimulation}, that adaptive timestepping integrators can still
be symplectic if they use a block-power-of-two scheme. Individual
timesteps (Section \ref{IndividualTimeSteps}) impose requirements that
reduce the symplecticity and angular momentum conservation from exact
to fifth-order (one order better than the trajectory error of our
integrator); nevertheless, we have shown experimentally in Sections
\ref{SymplecticityTests} and \ref{ManyBodySimulation} that the
benefits of fifth-order symplecticity and angular momentum
conservation are significant.

Though our code is not CPU-time competitive with standard
stellar-cluster simulations due to their use of the Ahmad-Cohen
nearest-neighbor scheme, we expect it will be useful in
direct-summation $N$-body simulations where phase space volume
conservation is more important than raw speed --- though it would be
interesting to see whether the Ahmad-Cohen scheme significantly
affects the symplecticity of our algorithm in practice, we have not
investigated this.  We intend to use our algorithm to check the
accuracy of dark matter halo evolution in larger, cosmological
$N$-body simulations, and to investigate the formation of caustics
in dark-matter distributions on a smaller scale.  In these
applications symplecticity is essential because the ``coldness'' of
the dark matter phase-space distribution is fundamental to the
dynamics.

Additionally, the algorithm was designed with the application to full
cosmological $N$-body simulations in mind.  The original motivation
for the work was to find a higher-order, symplectic algorithm which
takes only forward steps as an alternative to the leapfrog algorithm
used in cosmological simulations.  Standard compositional higher-order
symplectic integrators are not useful in this context because the
simulations of gas dynamics in high-accuracy cosmological simulations
become unstable under evolution backward in time, and these
integrators all have backwards timesteps.  We intend to explore this
application in a future paper.

\acknowledgements

We would like to thank Jack Wisdom and Gerry Sussman for helpful
discussions, encouragement, and inspiration.  We would also like to
thank an anonymous reviewer for helpful comments and for correcting an
error in Section \ref{VariationalIntegrators}.  This work was
supported by NSF grant AST-0407050 and NASA grant NNG06-GG99G.

\bibliography{ms}

\end{document}